\documentclass[Crown,sagev,times, doublespace]{sagej}

\usepackage{moreverb,url}
\usepackage{natbib}

\newcommand{\ds}{\displaystyle}
\def\volumeyear{2020}
\begin{document}
\runninghead{Tseng and Sim}
\title{Sample size planning for pilot studies}

\author{CHI-HONG TSENG\affilnum{1} , DANIELLE SIM\affilnum{1} }
\affiliation{\affilnum{1}Department of Medicine Statistics Core,
David Geffen School of Medicine,
University of California at Los Angeles,
Los Angeles, CA 90024}

\email{ctseng@mednet.ucla.edu}

\begin{abstract}

Pilot studies are often the first step of experimental research. It is usually on a smaller scale and the results can inform intervention development, study feasibility and how the study implementation will play out, if such a larger main study is undertaken. This paper illustrates the relationship between pilot study sample size and the performance study design of main studies. We present two simple sample size calculation methods to ensure adequate study planning for main studies. We use numerical examples and simulations to demonstrate the use and performance of proposed methods. Practical heuristic guidelines are provided based on the results. 

\end{abstract}

\maketitle

\section{Introduction}
Pilot studies are often conducted as the first step of experimental research. \citep{thabane2010, lancaster2004} It is usually on a smaller scale and the results can inform intervention development, study feasibility and how the study implementation will play out, if such a larger main study is undertaken. It also provides initial impression of treatment experience and the safety and efficacy profiles of treatments or interventions to help design the future main studies. \citep{thabane2010, lancaster2004, kistin2015} Thabane et al.\cite{thabane2010} provides an extensive definition of a pilot study and what it entails, as well as a checklist for elements to be included in the pilot study. 

Different areas of research have recognized the importance of pilot studies and have published their findings, explaining what the results of the pilot study mean in the context of their field, such as critical care and reproductive health \citep{arnold2009, van2002}. Other authors \citep{thabane2010, lancaster2004} have addressed the importance and strength that pilot studies provide for a larger randomized controlled trial (RCT). RCTs in its entirety are extensive, time consuming, and costly, and have the potential to unexpectedly fail when thorough preparations are not done. If pilot studies are done prior, there is a certain degree of preparation which can provide a foundation for the subsequent confirmatory studies. Lancaster et al.\cite{lancaster2004} and Arain et al.\cite{arain2010} have reviewed both publications and published pilot studies in major medical journals from 2000-2001 and 2007-2008 to observe which pilot studies were deemed successful for the main trial, to provide guidelines for researchers conducting a pilot study. These papers have been successful in highlighting the ‘what’ and ‘why’ of pilot studies, however, there is still a lack of research and guidelines in designing pilot studies and determining an appropriate sample size for pilot studies.

Various sample size recommendations have been proposed in response to the multipurpose nature of pilot studies.
Hertzog\cite{hertzog2008} proposed to determine sample size based on estimates' precision and illustrated the relationship between pilot study sample size and constructed confidence intervals to meet various goals of pilot studies. 
Cocks and Torgerson\cite{cocks2013} recommended a confidence interval approach in pilot studies to screen for intervention effects. \cite{julious2005} proposed a sample size of 12 per group for pilot studies based on feasibility, precision about the mean and variance, and regulatory considerations. Viechtbauer et al.\cite{viechtbauer2015} considered the sample size calculation of pilot studies with the goal of detecting potential problems aroused during the study. Other authors\cite{sim2012, whitehead2016, bell2018} advocated to estimate the pilot study sample size by minimizing the overall sample size needed for pilot and future studies together. 

One important aspect in research study design is to determine the sample size such that there is sufficient power in demonstrating meaningful effects to achieve study goals. The estimates from pilot studies provide useful information in calculating main studies' sample sizes. It is a common approach to use the observed effect size or outcome variability from pilot studies to facilitate sample size calculation for the main study. However, because pilot studies are inherently small, the estimates cannot be used with certainty. A small pilot study has a high potential to overestimate effect size or underestimate the variability, which can lead to underpowered study designs. Publications have suggested alternate methods to avoid this, such as using confidence intervals, as well as addressing other issues, such as skewed data. Some researchers\cite{browne1995, kieser1996, sim2012} recommended to use the upper confidence limit to reduce the chance of underestimating standard error in pilot studies when designing future studies. Teare et al.\cite{teare2014sample} used simulations to characterize the gain in estimation precision and power of main studies as sample size increases in pilot studies. Westlund and Stuart\cite{westlund2017} highlighted the challenge and insufficiency to use pilot study data to determine whether to conduct or to design a confirmatory study.

In this paper, we present the connection between pilot study sample size and the performance of the study design of main studies. Two simple sample size calculation methods are derived to ensure the adequate study planning of main studies, based on the estimated standard deviation and estimated effect size from pilot studies. We use numerical examples to demonstrate the implementation of these methods, and use simulations to confirm their performance. Practical heuristic guidelines are provided based on the results. 

\section{Methods}
\subsection{General framework for pilot study sample size calculation}
First, we describe the general framework that connects pilot study data with the study design and its performance in main studies. Let $X_p(N_p)$ be the collection of data from a pilot study of sample size $N_p$, and $\hat{\theta}_p$ be the estimator of population parameter $\theta$ based on $X_p(N_p)$. 
The distribution of $\hat{\theta}_p$ is a function of $\theta$ and $N_p$
\[
\hat{\theta}_p \sim F_{\theta_p}(.;\theta, N_p)
\]
When planning main studies, the sample size calculation is based on observed $\hat{\theta}_p$ from the pilot study and other parameter $\phi$, such that 
\[
N_m = G_{N_m}(\hat\theta_p) = \min\{N: \mbox{Prob}(T(X, N) \in C^T_{1-\alpha} | \hat{\theta}_p, \phi) > 1-\beta\}
\]
where $T(X, N)$ is the test statistic associated with the hypothesis of interest, and $T$ is a function of data $X$ and sample size $N$. $N_m$ is the calculated main study sample size, which is a function of $\hat{\theta_p}$, to have $1-\beta$ power with type I error $\alpha$. $C^T_{1-\alpha}$ is the critical region associated with the test statistic $T$ to reject the null hypothesis. 
However, the true power $Q$ for main studies is a function of $N_m$, 
\[
Q = G_Q(N_m) = \mbox{Prob}(T(X, N_m) \in C_{1-\alpha} | \theta, \phi)
\]
We can express the distribution of Q based on the distribution of $\hat{\theta}_p$, which is a function of $N_P$:
\begin{equation*}
  \begin{aligned}
F_Q(q) &= \mbox{Prob}(Q \leq q) = \mbox{Prob}(N_m \leq G_Q^{-1}(q)) \\
 &= 
\mbox{Prob}(\hat\theta_p \leq G_{N_m}^{-1} \circ G_Q^{-1}(q)) = F_{\theta_q}( G_{N_m}^{-1} \circ G_Q^{-1}(q) ; \theta, N_p)
  \end{aligned}
  \end{equation*}
We propose to estimate the pilot study sample size such that there is a less than $\tau$ chance that the true power of main studies is in the range $C_Q$
\[
N_p = \min\{N_p: \mbox{Prob}(Q \in C_Q) \leq \tau\}
\]
with $C_Q$ and $\tau$ are design parameters for pilot study sample size calculation.

\subsection{Sample size calculation based on estimated standard deviation from pilot studies}

In this section, we demonstrate our pilot sample size calculation method, when the outcome variability in pilot studies is incorporated in the sample size calculation of main studies.\citep{kistin2015}.  We first consider a one sample problem to compare the mean of a continuous variable to a practically meaningful value.  Let $Y_i$ be the outcome of interest for subject $i$, $i=1, \cdots, N_p$, in a pilot study of sample size $N_p$.  Let $\mu$ and $\sigma^2$ be the population mean and variance of variable $Y$. Pilot data sample mean ($\bar{Y}_p$)and variance ($S^2_p$) are consistent estimators for $\mu$ and $\sigma$, with $ \bar{Y}_p = \ds \frac{\sum_{i=1}^{N_p} Y_i}{N_p}$ and $S^2_p = \ds \frac{1}{N_p-1} \sum_{i=1}^{N_p} (Y_i-\bar{Y})^2$. With Normality assumption, the sample variance $S^2_p$ follows a Chi-square distribution with $N_p-1$ degrees of freedom;
\[
\frac{(N_p-1) S^2_p}{\sigma^2} = \frac{\sum_{i=1}^{N_p} (Y_i-\bar{Y})^2}{ \sigma^2} \sim \chi^2_{N_p-1}
\]
and
\[
\Pr( S^2_p < \frac{\sigma^2}{N_p-1} \chi^2_{q, N_p-1}) = q
\]
where $\chi^2_{q, N_p-1}$ is the 100$q$th percentile of $\chi^2$ distribution with $N_p-1$ degrees of freedom.

The hypothesis of interest for the main study is:
\[
H_0: \mu = 0 \quad \mbox{vs} \quad H_1: \mu > 0,
\]
With a practically meaningful effect $\delta$ and the sample variance $S^2_p$ from the pilot study, the sample size ($N_m$) required for the main study to have power of $1-\beta$ with type I error of $\alpha$ is
\[
N_m=(t_{1-\alpha/2} + t_{1-\beta})^2\frac{S^2_p}{\delta^2},
\]
where $t_{1-\alpha/2}$ and $t_{1-\beta}$ are the upper ${\alpha/2}$ and ${\beta}$ percentiles of $t$-distribution. Comparing to the calculated sample size based on true population variance $\sigma^2$,
\[
N_m^*=(t_{1-\alpha/2} + t_{1-\beta})^2\frac{\sigma^2}{\delta^2},
\]
$N_m$ will lead to an underpowered study design when $S^2_p$ underestimates true variance $\sigma^2$. 

Now we calculate the pilot study sample size such that main studies are more likely to have adequate power based on the pilot study estimate. This would require pilot study sample size ($N_p$) to be large enough that the chance underestimating $S^2_p$ is small:
\[
N_p= \min\{N: \Pr(S^2_p(N) < \sigma^2_L ) < p_0\},
\]
where $\sigma^2_L$ and $p_0$ are the design parameters of variance underpower threshold and underpower probability. 

The distribution of sample variance $S^2_p(N)$ is a function of sample size $N_p$. $\sigma^2_L$ is related to the corresponding power ($1-\beta_L$) of main studies, through main study sample size $N_L$: 
\[
N_L=(t_{1-\alpha/2} + t_{1-\beta})^2\frac{\sigma_L^2}{\delta^2}.
\]
The corresponding power of main studies is
\[
1-\beta_L = 1- T_{\delta\frac{ \sqrt{N_L}}{\sigma}}(t_{1-\alpha/2}).
\]
$T_\theta(\cdot)$ is the cumulative density function of non-central t-distribution with non centrality parameter $\theta$.

More generally, we can consider the sample size calculation of pilot studies to limit the chance to be underpowered or overpowered for main studies. Algorithm 1 below presents the steps to calculate sample size for pilot studies for one sample problem with continuous outcomes.

\begin{enumerate}
\item Identify study and design parameters:
\begin{itemize}
    \item $\sigma$: standard deviation of the outcome based on prior knowledge.  
    \item $\delta$: practically important study effect.
    \item $\alpha$ and $\beta$: type I error and type II error of the main study.
    \item $p_L$ and $\beta_L$: underpower probability and threshold: the main study aims to have less than $p_L$ chance with power lower than $1-\beta_L$.
    \item $p_U$ and $\beta_U$: overpower probability and threshold; the main study aims to have less than $p_U$ chance with power higher than power $1-\beta_U$.
\end{itemize}
\item Calculate the sample sizes of main studies corresponding to the underpower and overpower thresholds:
\[
    N_L=(t_{1-\alpha/2} + t_{1-\beta_L})^2\frac{\sigma^2}{\delta^2},
\]
\[
    N_U=(t_{1-\alpha/2} + t_{1-\beta_U})^2\frac{\sigma^2}{\delta^2},
\]
\item Solve for corresponding values of standard deviation:
\[
    \sigma^2_L=N_L \frac{\delta^2}{(t_{1-\alpha/2} + t_{1-\beta})^2},
\]
\[
    \sigma^2_U=N_U \frac{\delta^2}{(t_{1-\alpha/2} + t_{1-\beta})^2},
\]
\item Calculate the pilot study sample size based on the undepower and overpower probability with the $\chi^2$ distribution assumption of sample variance:
\[
N_{pL}= \min\{N: \Pr(S^2_p(N) < \sigma^2_L ) < p_L\}
\]
\[
N_{pU}= \min\{N: \Pr(S^2_p(N) > \sigma^2_U ) < p_U\}
\]
\item Calculate the final sample size.
\[
N_p=\max\{ N_{pL}, N_{pU} \}
\]
\end{enumerate}
For easy implementation, we can approximate step 4 calculation by \citep{canal2005}
\[
N^*_{pL}=2*\frac{z^2_{1-p_L}}{(\frac{\sigma^2_L}{\sigma^2}-1)^2}+1
\]
\[
N^*_{pU}=2*\frac{z^2_{1-p_U})^2}{(\frac{\sigma^2_U}{\sigma^2}-1)^2}+1
\]
\[
N^*_p=\max\{ N^*_{pL}, N^*_{pU} \}
\]

where $z_{1-p_L}$ and $z_{1-p_U}$ are the upper ${p_L}$ and ${p_U}$ percentiles of $z$-distribution.

We can extend this algorithm to two sample comparison problems, where the main study compares the mean of two populations and its sample size calculation is based on the (pooled) sample variance of pilot data, by changing the sample size calculation formula from one sample to two sample comparison in steps 2 and 3.

\subsection{Sample size calculation based on estimated effect size from pilot studies}
In this section we consider pilot studies to quantify the study effect for the sample size planning of main studies. With the same setting in the previous section, sample mean $ \bar{Y_p} = \ds \frac{\sum_{i=1}^{N_p} Y_i}{N_p}$ is a consistent estimator of $mu$ based on pilot data. Assuming Normality and known variance $\sigma^2$, 
the sample mean $\bar{Y_p}$ follows a Normal distribution: 
\[
\bar{Y_p} \sim N( \mu, \frac{\sigma^2}{N_p})
\]

The hypothesis of interest for the main study is :
\[
H_0: \mu = 0 \quad \mbox{vs} \quad H_1: \mu > 0.
\]
Using $\bar{Y_p}$ from the pilot study, the required sample size ($N_m^*$) for main studies to have power of $1-\beta$ with type I error of $\alpha$ is
\[
N_m=(t_{1-\alpha/2} + t_{1-\beta})^2\frac{\sigma^2}{\bar{Y_p}^2},
\]
where $t_{1-\alpha/2}$ and $t_{1-\beta}$ are the upper ${\alpha/2}$ and ${\beta}$ percentiles of $t$-distribution. Comparing to the calculated sample size based on the true population mean $\mu$,
\[
N_m^*=(t_{1-\alpha/2} + t_{1-\beta})^2\frac{\sigma^2}{\mu^2},
\]
$N_m$ will provide an underpowered study design if $\bar{Y_p}$ overestimates the true mean $\mu$.

Therefore the pilot study design would require the pilot study sample size ($N_p$) to be large enough, such that the chance to overestimate $\bar{Y_p}$ is limited:
\[
N_p= \min\{N: \Pr(\bar{Y_p}) > \mu_L ) < p_0\},
\]
where $\mu_L$ and $p_0$ are the design parameters of underpower threshold for mean and underpower probability. 

The distribution of sample mean $\bar{Y_p}$ is a function of sample size $N_p$. $\mu_L$ is related to the corresponding power ($1-\beta_L$) of the main study, through the calculated main study sample size $N_L$: 
\[
N_L=(t_{1-\alpha/2} + t_{1-\beta})^2\frac{\sigma^2}{\mu_L^2},
\],
and the corresponding power of the main study is
\[
1-\beta_L = 1- T_{\frac{\mu \sqrt{N_L}}{\sigma}}(t_{1-\alpha/2}).
\]
$T_\theta(\cdot)$ is the cumulative density function of non-central t-distribution with non centrality parameter $\theta$.

This method also applies when the goal is to quantify the effect size, which is the effect divided by standard deviation. The calculation using effect size with $\sigma=1$ and the calculation based on actual effect with known $\sigma$ are mathematically identical.

 Algorithm 2 below presents the steps to calculate sample size for pilot studies for one sample problem with continuous outcomes. The goal is to limit the chance of designing underpowered or overpowered main studies, based on the estimated study effect from pilot studies. 
 
\begin{enumerate}
\item Identify study and design parameters:
\begin{itemize}
    \item $\sigma$: standard deviation of the outcome.  
    \item $\mu_0$:  study effect based on prior knowledge.
    \item $\alpha$ and $\beta$: desired type I error and type II error of the main study
    \item $p_L$ and $\beta_L$: underpower probability and threshold; the main study aims to have less than $p_L$ chance with power lower than $1-\beta_L$.
    \item $p_U$ and $\beta_U$: overpower probability and threshold; the main study aims to have less than $p_U$ chance with power higher than power $1-\beta_U$.
\end{itemize}
\item Calculate the sample sizes of main studies corresponding the underpower and overpower thresholds:
\[
    N_L=(t_{1-\alpha/2} + t_{1-\beta_L})^2\frac{\sigma^2}{\mu_0^2},
\]
\[
    N_U=(t_{1-\alpha/2} + t_{1-\beta_U})^2\frac{\sigma^2}{\mu_0^2},
\]
\item Solve for corresponding values of study effect:
\[
    \mu_L=(t_{1-\alpha/2} + t_{1-\beta_L})\frac{\sigma}{\sqrt{N_L}},
\]
\[
    \mu_U=(t_{1-\alpha/2} + t_{1-\beta_U})^2\frac{\sigma}{\sqrt{N_U}},
\]
\item Calculate the pilot study sample size based underpower and overpower probability with Normality assumption of sample mean:
\[
N_{pL}= \min\{N: \Pr(\bar{Y_p}(N) > \mu_L ) < p_L\} = \frac{z^2_{1-p_L}\sigma^2}{(\mu_L-\mu_0)^2}
\]
\[
N_{pU}= \min\{N: \Pr(\bar{Y_p}(N) < \mu_U ) < p_U\} =
\frac{z^2_{1-p_U}\sigma^2}{(\mu_0-\mu_U)^2}
\]
\item Calculate the final sample size.
\[
N_p=\max\{ N_{pL}, N_{pU} \}
\]
\end{enumerate}

We can adjust this algorithm for two sample comparison problems, where the main study compares the mean of two populations and its sample size calculation is based on the estimated difference between groups from pilot data. This requires to change Steps 2, 3, and 4 according to the sample size calculation formula for two sample comparison problems.

\section{Numerical examples}
\subsection{Example 1, RMDQ example 1}

First we consider a low back pain example\cite{cocks2013}. The outcome measure is the Roland and Morris disability questionnaire(RMDQ), which has a standard deviation of 4. 
This is a two sample comparison problem with the main study as a randomized study comparing RMDQ between two groups of subjects receiving different treatments. The main study aims to detect practically meaningful effect of $delta = 1$ with $1-\beta = 80\%$ power at $\alpha = 5\%$ type I error rate. 
Now we follow modified algorithm 1 to design a pilot study such that the variance estimate will be reliable for the planning of main study in this study population. If the pilot study also includes subjects of two groups, we can estimate the variance based on pooled variance. 
\begin{itemize}
\item[1.] The goal is to design a pilot study such that there is less than $p_0 = 20\%$ chance that the calculated sample size for the main study would generate less than $1-\beta_L = 60\% $ power. 
\item[2.] With the standard deviation of $\sigma = 4$ based on prior knowledge, a sample of $N_L = 158$ subjects per group in the main study will have $1-\beta_L = 60\%$ power at the $\alpha = 5\%$ level to detect the clinically meaning effect of $\delta=1$ in RMDQ. 
\item[3.] The sample size of $N_L = 158$ subjects per group is also corresponding to $1-\beta = 80\%$ power if standard deviation is $\sigma_L = 3.16$. 
\item[4.] Pilot sample size of $N_pL = 12$ or larger will meet our goal that the sample standard deviation is smaller than $\sigma_L = 3.16$ with less than $p_0 = 20\%$ chance. 
\end{itemize}
Putting items 1-4 together, a pilot study with sample size of 12 will provide a reliable standard deviation estimate such that the study design of main studies will have less than $p_0 = 20\%$ chance to have power below $1-\beta_L = 60\% $. 

We further explore the pilot study sample size requirement to reliably quantify the standard deviation of the outcome in the study population with different parameters. Table 1 gives the pilot study sample size and its performance based on simulations, when the main study compares the mean of two groups. The range of underpower probability is from $p_0=$ 10\% to 30\% with $1-\beta_L=60\%$ power as the underpower threshold, initial guess of $\sigma$ is from 2 to 6, and practically meaning difference $\delta$ ranges from 1 to 4 between two groups. The left half of the table gives the derived pilot study sample sizes. The right half of the table gives underpower probability based on 1000 simulations. In the simulation, 1000 pilot study data are generated for each combination of design parameters, based on the sample size on the left size of the table. The pilot study data are generated based on Normal distribution, and the sample standard deviation is calculated. We use the 'pwr' package in R (www.r-project.org) to calculate the sample size of the main study and evaluate its power.

The table shows that, first, the required sample size increases as the underpower probability decreases. This is because a larger sample size provides more reliable standard deviation estimates to design the main study. Second, practically important effect $\delta$ has little impact on the required sample size. As the goal of pilot studies in this scenario is to quantify the standard deviation of outcome in the study population, we expect $\delta$ to have little impact on the sample size. Third, the required sample size is essentially invariant to input parameter $\sigma$ based on prior knowledge. This is a desirable quality because that often existing information on $\sigma$ is unavailable or unreliable, when conducting pilot studies. Fourth, the table suggests a useful heuristic guideline that pilot study sample sizes of 5, 12, and 25 correspond to 30\%, 20\% and 10\% underpower probability at 60\% power threshold in main studies. Last, the simulation study shows satisfactory performance of our method; the rounding error in the calculation is acceptable even when sample sizes are small. 

\begin{table}[h]  
\caption{Pilot sample sizes to quantify outcome variability. The left side of table gives pilot study sample sizes. The right side of the table gives underpower probabilities based on 1000 simulations. The underpower threshold is set at $60\%$ power.} 
\centering 
\begin{tabular}{l c rrrrr rrrrr} 
\hline\hline   
&\multicolumn{6}{c}{pilot study sample size} &\multicolumn{5}{c}{Simulation underpower probability}\\\hline
underpower & &\multicolumn{5}{c}{$\sigma$} &\multicolumn{5}{c}{$\sigma$}  \\ 
probability & {$\delta$} &  2 & 3 & 4 & 5 & 6 &  2 & 3 & 4 & 5 & 6\\
\hline
10\% & 1 &  25 & 25 &  25 &  25 &  25 & 8.7\% & 8.3\% & 9.1\% &6.9\% & 7.3\% \\
    & 2 & 25  & 25 &  25 &  25 &  25  & 4.3\% & 8.7\% &  7.3\% &7.2\% & 7.8\% \\
    & 3 &  24 &  25 &   25 & 25 &  25 & 4.0\% & 4.1\% & 6.5\% &7.9\% & 7.0\% \\
    & 4 &  24 &  25 &  25 & 25 &  25 &  1.2\% & 4.5\% & 4.2\% &6.8\% & 6.5\% \\
    \hline
20\% & 1 & 12 &  12 &  12 &  12 &  12 & 16.5\% & 22.1\% & 18.0\% & 17.2\% & 19.5\% \\
    &  2&  11 &  12 &  12 &  12 &  12 & 13.9\% & 19.4\% & 20.2\% & 21.1\% & 19.8\% \\
    & 3 &   11 &  11 &  12 &  12 &   12 & 14.8\% & 13.9\% & 16.9\% & 19.6\% & 18.5\% \\
    & 4 &    11 &  11 &  11 &  12 &  12& 7.2\% & 14.2\% & 15.5\% & 16.7\% & 19.8\% \\
 \hline
30\% & 1 &  5 & 5 &  5  &  5  &  5 & 34.0\% & 37.9\% & 32.9\% & 34.4\% & 36.9\% \\
    & 2 &  5  & 5 &  5 &  5  &  5 & 32.4\% & 35.9\% & 39.6\% & 35.3\% & 36.2\% \\
    & 3 &  5  & 5 &  5 &  5  &  5 & 29.1\% & 29.0\% & 35.7\% & 35.0\% & 35.5\% \\
    &4  &  5 & 5 &   5 &  5  &  5 & 34.0\% & 29.7\% & 30.0\% & 33.7\% & 36.8\% \\
\hline\hline 
\end{tabular}
\end{table}

\subsection{Example 2, RMDQ example 2}
Next, we consider the situation to use pilot studies to quantify the study effect. Both the pilot study and the main study compare RMDQ between two treatment groups of equal sample size. We also assume a medium effect size of 0.5 ($\mu_0=2, \sigma=4$) between treatment groups. The pilot study sample size calculation will use  algorithm 2, with modification to two sample problems, to obtain reliable effect size between the treatments.

\begin{itemize}
\item[1.] The goal is to design a pilot study such that there is less than $p_0 = 30\%$ chance that the calculated sample size for the main study would generate less than $1-\beta_L = 60\% $ power. 
\item[2.] With the standard deviation of $\sigma = 4$ and $\mu_0=2$ based on prior knowledge, a sample of $N_L = 40$ subjects per group in the main study will have $1-\beta_L = 60\%$ power at the $\alpha = 5\%$ level to detect effect size of 0.5. 
\item[3.] The sample size of $N_L = 40$ subjects per group is also corresponding to $1-\beta = 80\%$ power if the effect size is 0.63 in the main study. 
\item[4.] Pilot sample size of $N_pL = 32$ or larger per group will meet our goal that the estimated effect size is greater than 0.63 with less than $p_0 = 30\%$ chance. 
\end{itemize}
Therefore, a pilot study with sample size of 32 per group will provide a reliable estimate for the study effect such that the study design of main studies will have less than $p_0 = 30\%$ chance to have less than $1-\beta_L = 60\% $ power. 

Based on the same two group comparison setting, table 2 gives the pilot study sample size and its performance based on simulation, with underpower probability $p_0=$ 20\% to 40\% at $1-\beta_L=60\%$ power threshold . The left side of the table gives the derived pilot study sample size per group. The right side of the table gives underpower probability based on 1000 simulations. On the last row of the table, we give the main study sample size, with given effect size $\mu_0 =$ 0.2, 0.5, and 0.8, which corresponds to small, medium and large effect (with $\sigma=1)$. In the simulation, 1000 pilot study data are generated for each combination of design parameters, based on the sample size on the left size of the table. We generate Normally distributed pilot data, estimate effect size, and then use the 'pwr' package in R (www.r-project.org) to calculate the sample size of the main study and evaluate the power.

As expected, the larger sample size of pilot study providing better precision on the effect size estimate, results in smaller underpower probability. The table also shows that the effect size $\mu_0$ has large impact on the pilot size sample. When the effect size is small, one needs large sample size in pilot studies to quantify the treatment effect. This sample size calculation can only work reliably when some knowledge of effect size is available, as the effect size has large impact on the needed sample size. With underpower probabilities of 30\% and 35\%, the required pilot study sample size is about 50\% and 25\% of the sample size of main studies. These numbers may be useful as a heuristic guideline for pilot sample size calculation. The pilot sample size can exceed the sample size required in the main study, when requiring underpower probabilities less than $25\%$. This highlights the difficulty in using small pilot study to quantify the effect size to determine the design of main studies. Simulation study suggests satisfactory performance of our method; the rounding error in the calculation is acceptable even when sample size is small. 

\begin{table}[h]  
\caption{Pilot study sample size to quantify study effect size. The left side of the table gives pilot study sample sizes. The right side of the table gives underpower probabilities based on 1000 simulations. The underpower threshold is set at $60\%$ power.} 
\centering 
\begin{tabular}{lrrrrrrr} 
\hline\hline   
\multicolumn{4}{c}{pilot data sample size} & \multicolumn{4}{c}{simulation underpower probability} \\\hline
underpower &\multicolumn{3}{c}{effect size} & &\multicolumn{3}{c}{effect size} \\
probability &  0.2 & 0.5 &0.8 &&  0.2 & 0.5 &0.8 \\
\hline
20\% & 501 & 81 & 32 &\quad& 19.2\% & 20.3\% & 20.4\% \\
25\%  & 322 & 52 & 21 &\quad&25.0\% & 25.4\% & 28.1\% \\
30\% & 195 & 32 & 13  &\quad& 32.0\% & 30.8\% & 33.5\% \\
35\% & 106 & 17& 7    &\quad&34.7\% & 37.5\% & 36.5\% \\
40\% & 46 & 8 & 3     &\quad& 42.8\% & 42.5\% & 48.7\% \\
\hline
main study sample size & 394 & 64 & 26 && & &\\
\hline\hline 
\end{tabular}
\end{table}
\subsection{Example 3, Fall prevention example}

Considering a fall prevention example\citep{cocks2013}, the goal is to demonstrate the reduction of the proportion of older people who are at risk of falling from 50\% to 40\% with intervention. Both the pilot study and the main study consist of two treatment groups of equal sample size. The main study aims to demonstrate the impact of intervention on the reduction of proportion of subjects at risk of falling($p_{fall}$) with $1-\beta = 80\%$ power at the $\alpha = 5\%$ type I error rate. The hypothesis is:
\[
H_0: p_{fall} = 50\% \quad \mbox{vs.} \quad H_1: p_{fall} < 50\%.
\]
The pilot study goal is to quantify the intervention effect.

We follow the modified algorithm 2 and use Arcsine transformation to calculate effect sizes.

\begin{itemize}
\item[1.] The goal is to design a pilot study such that there is less than $p_0 = 30\%$ chance that the calculated sample size for main studies would generate less than $1-\beta_L = 60\% $ power. 
\item[2.] Based on Arcsine transformation, the intervention effect of reduction of the proportion of older people who are at risk of falling from 50\% to 40\% is equivalent to $\mu_0= 0.20$ and $\sigma = 1$. 
\[
\mu_0 = 2 \mbox{arcsin}(\sqrt{0.5}) - 2\mbox{arcsin}(\sqrt{0.4}) = 0.20, \quad \sigma=1
\]
A sample of $N_L = 246$ subjects per group in the main study will provide $1-\beta_L = 60\%$ power at the $\alpha = 5\%$ level to detect effect size $\mu_0= 0.20$. 
\item[3.] The sample size of $N_L = 246$ per group subjects in the main study is also corresponding to $1-\beta = 80\%$ power if the effect size $\mu_L = 0.253$. 
\item[4.] Pilot sample size of $N_pL = 195$ per group or larger will meet our goal that the estimated effect size is greater than $\mu_L = 0.253$ with less than $p_0 = 30\%$ chance. 
\end{itemize}
Based on above calculation, a pilot study with sample size of 195 per group is needed to reliably quantify the intervention effect for the study design of main studies. This pilot study sample size is about half of the sample size needed in the main study; a sample of 394 subjects per group in the main study have 80\% power at the 5\% level to demonstrate the reduction on the proportion of older people who are at risk of falling from 50\% to 40\%.

\section{Conclusion}

This paper illustrates the relationship between the pilot study sample size and its impact on study design of the main study. We present a general framework to calculation pilot study sample size to ensure the adequate study design of main studies. We first use the standard deviation from pilot data as the basis of sample size calculation for main studies. This method suggests that a pilot study with sample sizes of 5, 12, and 25 correspond to 30\%, 20\% and 10\% underpower probability in the main study, if setting 60\% power as the underpower threshold. The pilot study sample size appears insensitive to true standard deviation and practical important difference in the calculation. This is a desirable quality as the true standard deviation is often unknown at the time of pilot study. The second method uses the observed effect size from the pilot data to calculate the main study sample size. Many authors\citep{westlund2017, sim2019} have documented the problems associated with this method. Our results agree. First, the calculated sample size is sensitive to true effect size, which is usually unknown to begin with. Second, to maintain a low underpower probability in main studies requires a very large sample size in pilot studies, even when the true effect size is specified correctly. Our results also suggests a heuristic rule for this method: if pilot study sample size is about 50\% and 25\% of the sample size of the main study, the corresponding underpower probability is about 30\% and 35\%, with 60\% power as the underpower threshold. 

Our methods have limitations. First, our closed form calculation assumes Normality. This assumption can result in large bias when the outcome variable is skewed and the pilot study sample size is small.\citep{lumley2002} Using simulation to complete some of the steps may be helpful in this situation. We present the algorithms with one sample and two sample problems of equal sample size with continuous outcomes. One can extend it to binary outcomes based on Arcsine transformation, as we illustrate in example 3. Extensions to unequal sample sizes, survival outcomes or equivalence/non-inferiority studies should be feasible with some modification. In the problems where closed form sample size calculation expressions are not available\cite{li2018}, some of the steps may be achievable with simulations. Other extensions, such as to use pilot studies to screen for best treatment\citep{simon1985}, to include multiple outcomes\citep{jung2005}, or to incorporate interim analysis to stop early for futility\citep{simon1989}, are interesting future work. 

\section{Supplemental materials}
R codes for one sample and two sample pilot study sample size calculation are included.

\end{document}